\documentclass[12pt,preprint]{aastex}

\begin{document}


\title{Static and Impulsive Models of Solar Active Regions}
\author{S. Patsourakos\footnote{also Center for Earth Observing and Space Research, Institute for Computational 
Sciences - College of Science,
George Mason University, Fairfax, VA 22030}
\,and J. A. Klimchuk
\footnote{Current address:  NASA Goddard Space Flight Center, Code 671, 
Greenbelt, MD  20771}}
\affil{Naval Research Laboratory, Space Science Division,
Washington,  DC 20375 \\
patsourakos@nrl.navy.mil}



\begin{abstract}

The physical modeling of  active regions (ARs) and 
of the global coronal   
is receiving increasing interest lately. Recent attempts to model ARs  
using static equilibrium models were quite successful in
reproducing AR images of hot soft X-ray (SXR) loops.
They however failed to predict the bright EUV warm loops
permeating ARs: the synthetic images were dominated by 
intense footpoint emission. 
We demonstrate that this failure
is due to  the very weak dependence
of loop temperature on loop length which cannot simultaneously
account for both hot and warm loops in the same AR.
We then consider time-dependent
AR models based on nanoflare heating. We demonstrate that such
models can simultaneously reproduce  EUV and SXR loops in ARs.
Moreover, they predict radial intensity variations consistent with the 
localized core and extended emissions in SXR and EUV AR observations
respectively. We finally show how the AR morphology 
can be used as a gauge of the  	properties 
(duration, energy, spatial dependence, repetition time)
of the  impulsive heating.

\end{abstract}


\keywords{hydrodynamics; Sun : Coronal Heating, Sun : Corona}


\section{Introduction}

Modeling ensembles of coronal loops in active regions or over the  full
Sun is a rapidly emerging new field in the study of the solar
corona. Such studies attempt to reproduce  generic properties
of ARs and of the  global corona such as space  integrated intensities
and overall morphology. This allows for the determination various  properties of coronal heating
such as its dependence on loop length and magnetic field 
thereby testing  coronal heating mechanisms.  
Moreover such studies eliminate potential
selection effects which may enter into  studies of individual loops.
Finally, AR and global coronal models
pave the way for the construction of physics-based models
of  the EUV and SXR irradiance, an important contributor
to Space Weather conditions.

The majority of the attempted AR and global coronal
models  were based on static equilibrium coronal loop
models produced with steady heating (Schrijver et al. 2004; 
Mok et al. 2005; 
Warren $\&$ Winebarger 2006; Lundquist,  Fisher $\&$ McTiernan 2007).
They were quite successful at reproducing
the general appearance of the corona in hot emissions
($>$ 3 MK) observed in the SXR by the Soft X-ray Telescope (SXT) 
on {\it Yohkoh}.
However, synthetic EUV images from the static models 
did not show evidence of the warm loops seen in the real observations, but 
instead were 
dominated by intense footpoint (moss)
emission.  
Steady heating that is highly concentrated near the coronal base can produce 
time dependent behavior called thermal non-equilibrium.  This gives rise to transient 
EUV loops (Mok et al. 2008; 
Klimchuk, Karpen, \& Patsourakos 2007), but whether their properties are fully 
consistent with observations has yet to be determined.

Most theories of coronal heating predict that energy is released
impulsively on individual flux strands (Klimchuk 2006).  This includes
both AC (wave) and DC (reconnection-type) heating.  
Following
this paradigm coronal loops are viewed as ensembles
of unresolved, impulsively heated strands.
A very promising
idea first proposed by Parker (1988) is that the coronal field becomes
tangled on small scales due to the random footpoint motions associated
with turbulent photospheric convection.  Current sheets develop at the
interfaces between individual misaligned strands, and it has recently
been shown that an explosive instability called the secondary
instability occurs when the misalignment angle reaches a critical value
(Dahlburg, Klimchuk $\&$ Antiochos 2005).  This may be the physical nature of the
nanoflares postulated by Parker.  The heating function we have used in
our simulations is appropriate to nanoflares that are initiated at a
critical angle.  We note that the upward Poynting flux associated with
observed photospheric field strengths and observed photospheric
velocities is consistent with coronal heating requirements
(e.g. Abramenko, Pevtsov $\&$ Romano 2006), but the
efficiency of field line tangling is only now being addressed
quantitatively with high resolution magnetogram movies from Hinode.  We
also note that direct observations of the kinds of nanoflares we are
discussing are scarce (Katsukawa $\&$ Tsuneta  2001).
The small distinct brightenings that are sometimes called nanoflares are
much different from the unresolved energy releases on long field lines
that we consider here. 

But why do static equilibrium models fail to reproduce
the coronal emission patterns in the EUV? Can impulsive
heating models do any better? And how can AR morphology 
be used to infer the properties of impulsive heating?
With this paper we address these important questions.

\section{AR Simulations}

For the calculations reported in this paper
we use the  0D  hydrodynamic loop model we call
enthalpy-based thermal evolution of loops (EBTEL), described 
in Klimchuk, Patsourakos $\&$ Cargill (2008).
For a given temporal profile of the heating
EBTEL calculates the temporal evolution
of the spatially averaged temperature, density, and pressure 
along the loop.  It also provides the Differential Emission Measure 
distribution (DEM(T)) for both the 
coronal and footpoint (i.e., transition region) sections of the loop. 
For steady heating,  static equilibrium solutions
are calculated. EBTEL is capable of relatively accurately
mimicking complex 1D hydrodynamic simulations, with however
much less demands in terms of CPU time, which makes it
particularly useful for parametric investigations of multitudes of loops.
We calculated hydrodynamic models for 
26 loops with lengths in the range 
50-150 Mm, typical of observed AR loops.
For the construction of the AR images
we assumed that the loops have semi-circular shapes
and are nested the one on top of the other, forming
an arcade which emulates the simplest form
of AR topology.  

\subsection{Static Models}
We first calculated a static equilibrium AR model. 
The steady volumetric  heating  $H$ supplied to   
a loop   with  length  $L$ was : 
\begin{equation}
H=H_{0}{(L/L_{0})}^{\alpha},
\end{equation} 
with $H_{0}$ the heating magnitude,  
$L_{0}$ the  length of the shortest 
loop  and $\alpha$ a scaling-law index
which depends on the details of the specific
coronal heating mechanism (e.g., Mandrini et al. 2000).  We chose 
$L_0=50$ Mm,   
$H_{0}$=0.01 \, $\mathrm{erg\,{cm}^{-3}{s}^{-1}}$,
and   $\alpha=-2.8$. This particular $\alpha$
corresponds  to heating  associated with the tangling of the magnetic field  
by photospheric convection.  A nanoflare occurs when the misalignment  
between adjacent flux strands reaches a critical angle.  Similar $\alpha$ values
were found to provide the best match between static models 
and AR and full Sun SXR images (Schrijver et al. 2002; Warren $\&$ Winebarger 2006;  Lundquist,  Fisher $\&$ 
McTiernan 2008). In concert with these studies, 
$H_{0}$ was chosen so that  the temperature of the shortest and consequently
hottest (according to Equation 1) loop of the  arcade was $\approx$ 4 MK,
consistent with the temperature of bright SXR loops
in AR cores.   The $DEM$ distributions  for the coronal and footpoint sections
of each simulated loop were folded with the temperature response functions of the 171 \AA \,
and AlMg channels of TRACE and SXT respectively  to determine the 
corresponding intensities. For the calculation
of the intensities we assumed that simulated loops 
had a diameter of 3 Mm, consistent with
typical  widths of observed loops.

In Figure \ref{fig:int_base} (left column)  we plot the variation of   intensity  
across the loop arcade while  in Figure \ref{fig:im_base}   (left column)  we display 
the corresponding synthetic images. For building the images we assume that 
the AR is viewed face-on and  for clarity reasons we display every fourth loop of our arcade.
We applied  convenient box-cars to the images to emulate the different spatial resolution
of SXT ($\approx $ 5 arcsec) and TRACE ($\approx$ 1 arcsec).
We assumed that the coronal emission is distributed uniformly along the loop.  This 
is reasonable because our loops are shorter than a gravitational 
scale height, so there will be minimal gravitational stratification. Furthermore, 
the coronal temperature varies by only about  
$50\%$ along most of the length of equilibrium loops (Klimchuk et al. 2008).  
Temperature and density of course vary dramatically in the transition region, 
but the thickness of transition region is generally unresolvable, so we spread the 
emission over 2 Mm for convenience and clarity. The magnitude of the integrated 
transition region emission is correct.

From Figure  \ref{fig:int_base} we note that the SXT emission is considerably weaker  
at the footpoints than in the corona.  
This is because the footpoint temperatures are generally below 2 MK, where SXT has 
greatly reduced sensitivity.
Furthermore, from  
Figure \ref{fig:im_base} the synthetic TRACE image is completely dominated by 
the footoints.  There is little evidence of EUV loops, which is at 
odds with the multitudes of EUV loops seen in the majority of observed ARs.
Our results are broadly consistent with previous studies employing static heating.

Why is it that static equilibrium models fail to predict bright EUV loops 
together with bright SXR loops? 
This "pathology"
is related to the fact that under static equilibrium
conditions the loop temperature has a very weak dependence
on the loop length. For instance, the Rosner et al. (1978) scaling law
predicts that the apex temperature is related to the 
heating rate and length according to $T_a \propto {H}^{2/7} L^{4/7}$.  
Therefore, for the employed value $\alpha=-2.8$
we have from Equation 1  that
$T_{a} \propto L^{-0.22}$. This means that  the 
temperature is reduced by a factor of only 
$3^{-0.22}=0.8$ in going 
from the shortest loop (4 MK) to the longest loop (3.1 MK) in the arcade.  
As shown Figure \ref{fig:temp_base}, 
our simulations closely follow the above scaling law.
Therefore, none of the 
loops  has 1-2 MK plasma in the coronal section, which is necessary to produce 
significant EUV emission. All of this plasma resides at the footpoints, which 
is where the strong TRACE emission originates.
Obviously, we could decrease $H_{0}$ to produce strong TRACE emission in the 
corona, but then the SXT emission would be dramatically reduced.  The only 
way to have both bright TRACE loops and bright SXT loops in the same arcade 
is for the heating rate to have 
a much stronger dependence 
on loop length.  Our example arcade requires $\alpha \simeq -6$.  We are not 
aware of any coronal heating
mechanism  with such an extreme dependence on $L$ (e.g. Mandrini et al. (2001)).

\subsection{Impulsive Models}
\begin{figure}[!h]
\plotone{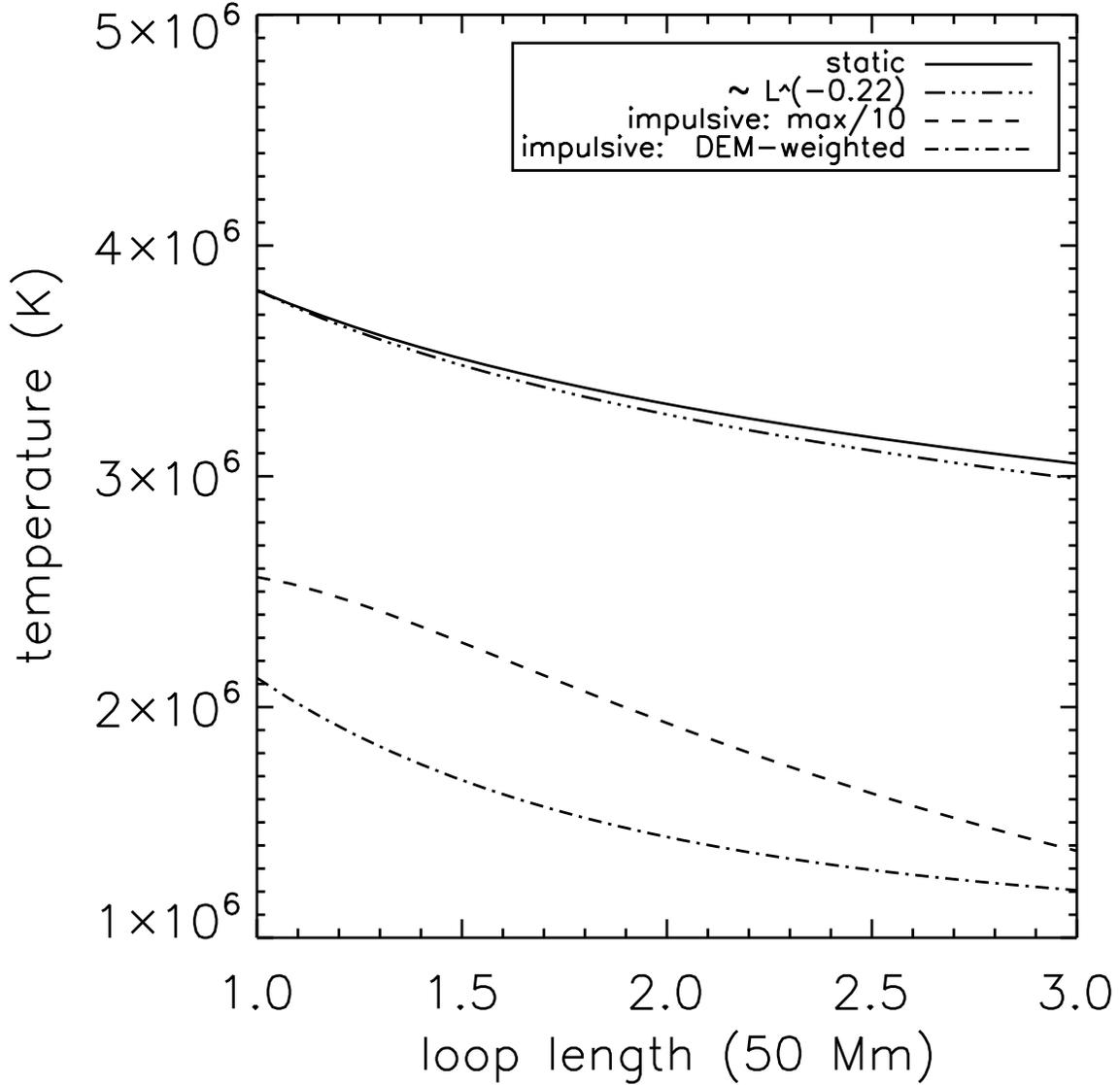}
\caption{Variation of the temperature across the loop arcade.  
Static equilibrium  model (solid line)
and fit $\propto$ $L^{-0.22}$ (dash-triple dot). 
Impulsively  heated model: 
maximum temperature divided by 10 (dashed line); $DEM$-weighted mean temperature
(dashed-dotted line).}
\label{fig:temp_base}
\end{figure}

We then considered models with 
impulsive heating, using  the  same 
loop  arcade of the previous section. 
 We started with static 
equilibria having an average coronal temperature
near 0.5 MK.  We heated the loops with a triangular pulse 
lasting  50 s and let the loops cool for 8500 s, by which time the 
temperature had cooled below 1 MK. The amplitude of the 
heat pulses varied from loop to loop according to Equation 1, with 
$\alpha = -2.8$, as before, and 
$H_{0}=2.5 \, \mathrm{erg\,{cm}^{-3}\,{s}^{-1}}$.  This produced an 
average $DEM$ for the arcade that peaks near 2.0  MK, similar to that of observed 
active regions (e.g., Brosius et al. 1996). 
Each loop reaches a maximum temperature exceeding 30 MK 
(Figure \ref{fig:temp_base}), but this happens  
very early in the heating event, when the density is very low.  The 
$DEM$-weighted mean temperature $T_{DEM}$ is only 1-2 MK.

\begin{figure}[!h]
\plotone{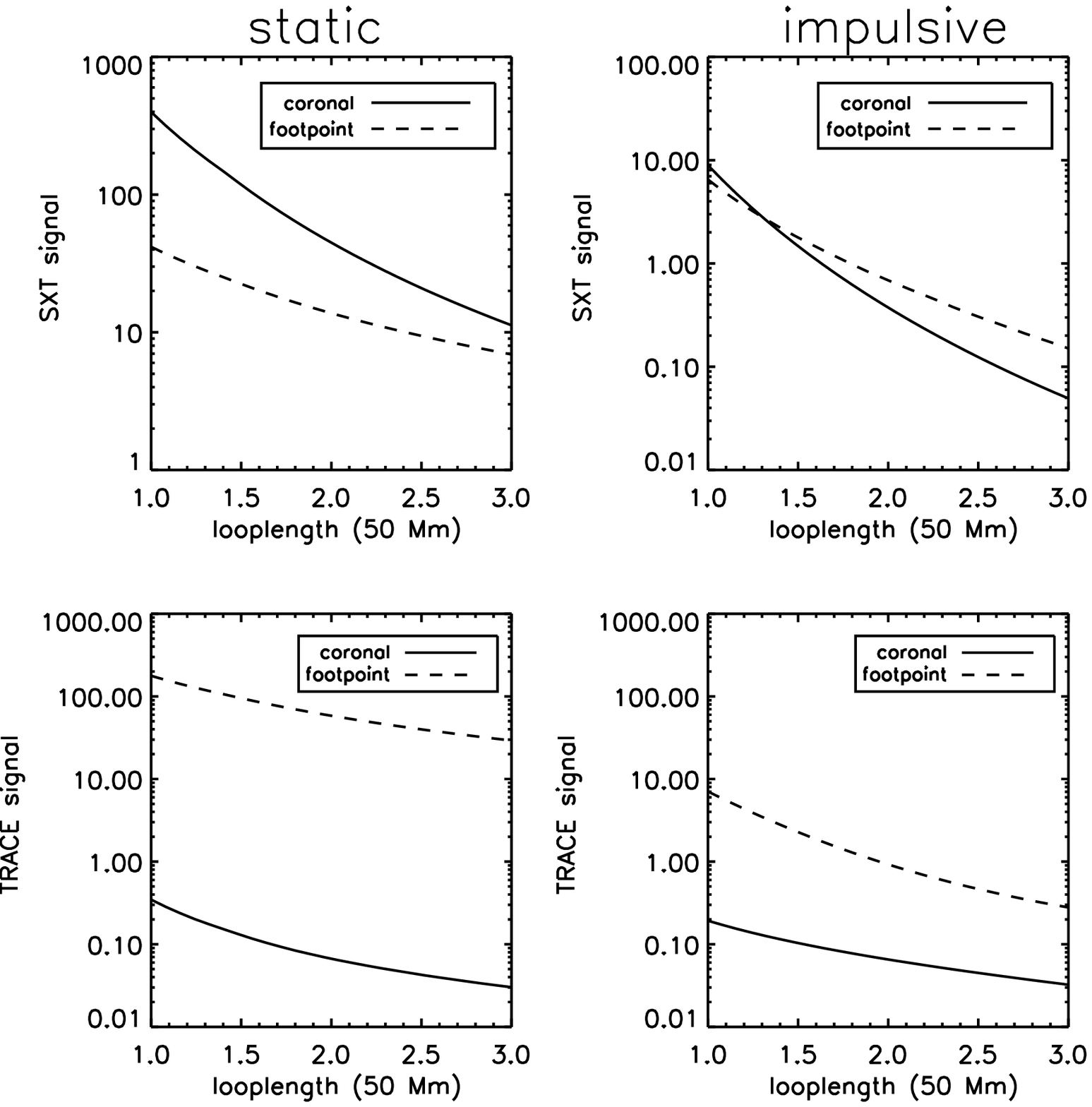}
\caption{Variation of the intensity across the loop arcade for static equilibrium
(left column)
and for impulsive heating (right column) for
SXT and TRACE. We plot
the intensities for both the coronal (solid line) and transition region section
(dashed line) of the loops. For impulsive heating the time-averaged intensities
are plotted. Intensities are in units of  DN/pix/s.}
\label{fig:int_base}
\end{figure}

\begin{figure}[!h]
\plotone{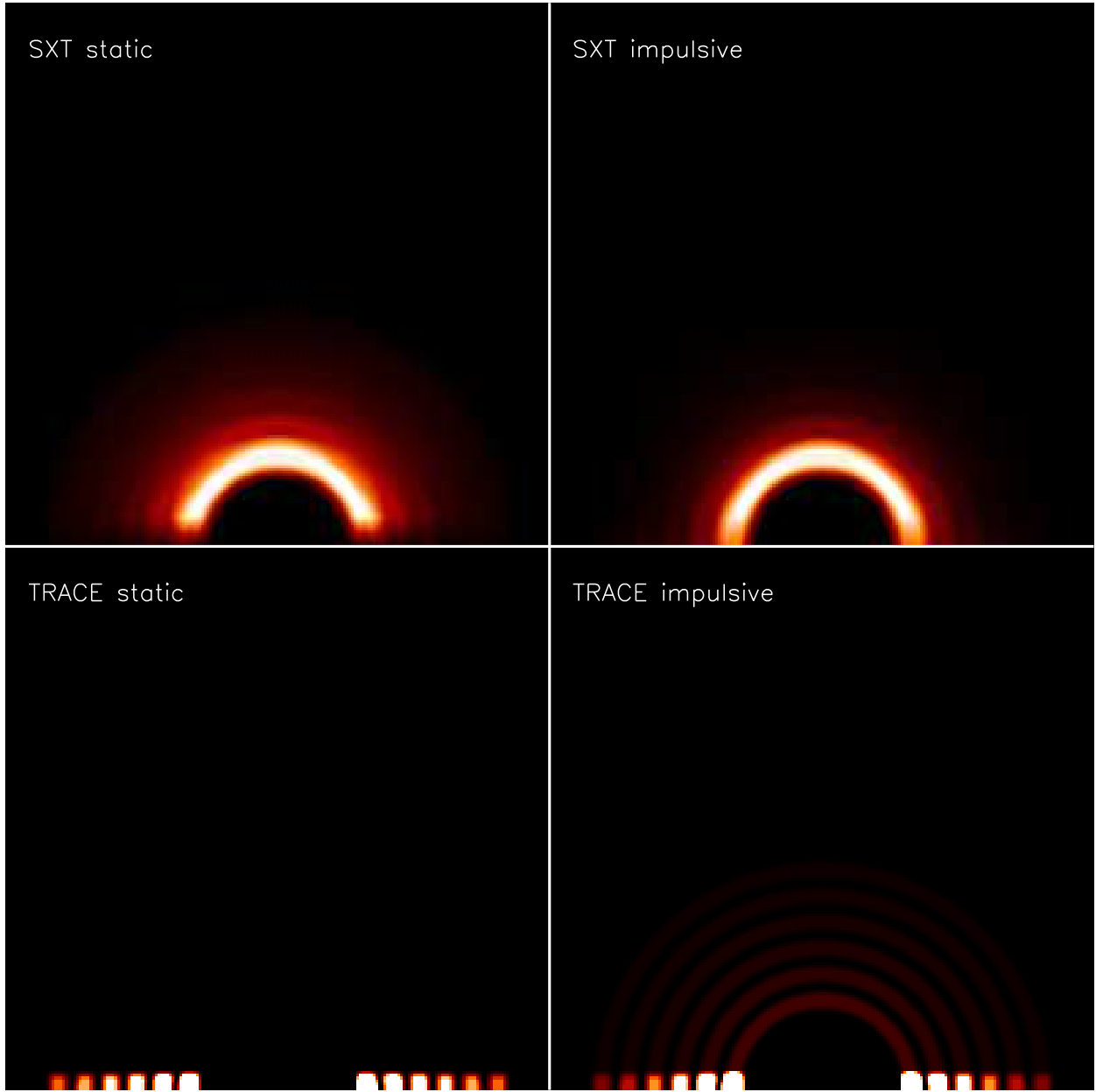}
\caption{SXT and TRACE images of the simulated AR loop arcade for static equilibrium
(left column) and impulsive heating (right column). The images for the impulsive heating 
correspond to a time-average over the corresponding simulations. It is assumed
that the AR is observed face-on. The images are smoothed
with boxcars  consistent with the  instrument resolution.
The AR baseline has  a length of $\approx$ 100 Mm. Intensity represented with color
increasing from black to red to white. Each image is normalized individually.}
\label{fig:im_base}
\end{figure}

We determined temporally averaged 
TRACE and SXT intensities for 
each loop simulation.  
Time averaging over the duration of a  loop simulation is equivalent to
taking a snapshot of a loop containing
a large number of impulsively heated  
unresolved strands at different stages of heating and cooling.  As before, 
we produced synthetic images of the arcade by assuming that the 
time-averaged intensities are uniform along each 
the loop.  Our 1D hydro simulations indicate that this is a reasonable  
approximation (e.g., Klimchuk et al. 2006).
The right columns of 
Figures \ref{fig:int_base} and \ref{fig:im_base} show the  
intensity variation across the arcade and the synthetic image, 
respectively.  


We  note in Figure \ref{fig:int_base}   that the TRACE emission
from the footpoints is a factor 
of $\approx$ 3-100 smaller for impulsive heating than for static heating, 
wheres the coronal emission is about the same in the two cases.  The brightness 
contrast between the footpoints and corona is therefore significantly reduced for impulsive 
heating.  This leads to a TRACE image in which both the coronal and footpoint emissions 
can be readily discerned   
(Figure $\ref{fig:im_base}$ right panel).  This is not true for the static model 
(left panel), where the coronal emission is overwhelmed by the footpoint emission. 
The footpoint to
corona intensity ratios $I_{foot}/I_{corona}$ are of order 10 in the impulsive 
model and 1000 in the static model.  Observed values are in the range
$\approx$ 2-20.   
The smaller observed ratios could be due to spicular absorption
of the footpoint emission (e.g., Daw,  Deluca $\&$   Golub 1995;  De Pontieu et al. 1999).
Using the analytical expressions of Anzer $\&$ Heinzel (2005)  for absorption at TRACE wavelengths 
and typical physical
parameters of spicules given in Table-1 of  Tsiropoula $\&$ Schmieder (1993)
we found that attenuation factors of the 171 TRACE emission of about 10 can
be achieved. As a matter of fact a recent SUMER/EIS study of moss intensities formed
above and below the head of the hydrogen Lyman continuum at 912 \AA \,
demonstrated that indeed sizeble absorption occurs over moss regions (De Pontieu et al. 2008).   
Future AR  models would probably need to incorporate
absorption effects.
Warren and Winebarger (2007) modeled an observed AR using 1D simulations and 
also found that impulsive heating increases the visibility of EUV loops compared 
to static equilibrium.

It can also be seen in Figure \ref{fig:int_base} that brightness of the corona in TRACE 
relative to SXT is roughly 
two orders of magnitude larger in the impulsive model than in the static model.  
This suggest that bright TRACE loops are much more likely to be seen together with 
bright SXT loops in the same active region if the heating is impulsive. 

The increased TRACE-to-SXT coronal intensity ratio and the reduced 
footpoint-to-corona TRACE contrast can both be understood as follows.  
In static equilibrium, the optimum 
temperature for a particular waveband occurs either in the corona 
or at the footpoints, but not at both locations.  For apex temperatures $>$ 2 MK, the coronal 
plasma is too hot to emit appreciable TRACE emission, and only the footpoints are bright.  
This is the case for all of the loops in the static arcade. Impulsively heated loops are fundamentally different in that they experience a wide range of 
coronal and footpoint temperatures over the course of their evolution.  When an 
impulsively heated loop starts to cool, TRACE emission occurs first at the footpoints 
and then over the full length of the 
loop when the coronal temperature drops below 2 MK.  A bundle of impulsively heated 
strands will therefore have significant TRACE emission both in the corona and at the footpoints. 

A final interesting property revealed in Figure \ref{fig:int_base} is the distribution 
of coronal emission across the arcade.  SXT intensities decrease rapidly with loop 
length, especially in the impulsive model, while TRACE intensities decrease much more 
slowly.  As a consequence, the SXT emission is concentrated in the core of the arcade, and  
the TRACE emission is much more extended (Figure $\ref{fig:im_base}$).  This agrees well 
with AR observations.
The strong decrease in the SXR emission with increasing $L$ is because  
$T_{DEM}$, the dominant temperature of the plasma, drops below 2 MK for the longer
loops (Figure \ref{fig:temp_base}). The sensitivity of SXT is a rapidly 
decreasing function of temperature in this range.  
On the other hand, the $T_{DEM}$ values are in the range where TRACE has good sensitivity, so 
the TRACE intensity gradient across the arcade is more shallow.

\begin{figure}[!h]
\plotone{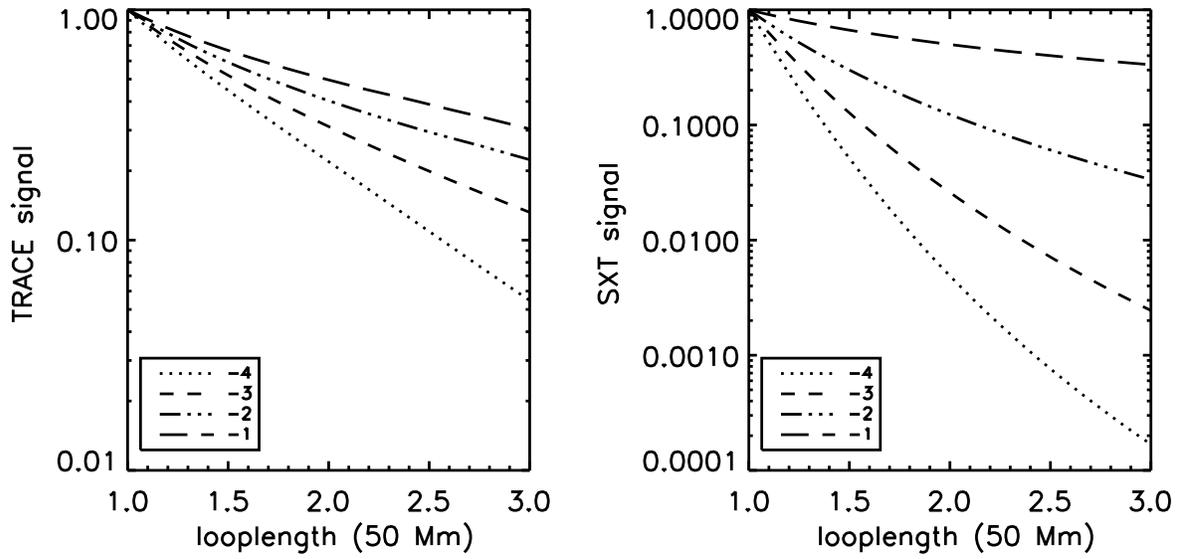}
\caption{Variation of the normalized coronal intensity across the loop arcade for
different values of the power-law index $\alpha$, determining how the nanoflare
energy is distributed across loops with different $L$, for TRACE (left panel) and 
SXT (right panel).}
\label{fig:paramet}
\end{figure}

We then examined how the AR morphology depends on the   properties
of the impulsive heating.
We first considered the effect of varying the index $\alpha$ of Equation 1 
by taking the following values: -4, -3, -2, -1.
All other aspects of the simulation were the same as 
for the $\alpha = -2.8$ base simulation discussed above. The resulting intensity
variations across the loop arcade  for TRACE and SXT are given in
Figure \ref{fig:paramet}. Not surprisingly, we found that a
stronger dependence of impulsive heating on $L$ (i.e.
a more negative $\alpha$) produces a steeper intensity
drop-off across the arcade. 
Note that that if we were to observe our 
arcade off-limb,  the model with $\alpha$=-1 would   
yield  an intensity  scale-height consistent with off-limb observations
of the unresolved EUV corona (Cirtain et al. 2005).

\begin{figure}[!h]
\plotone{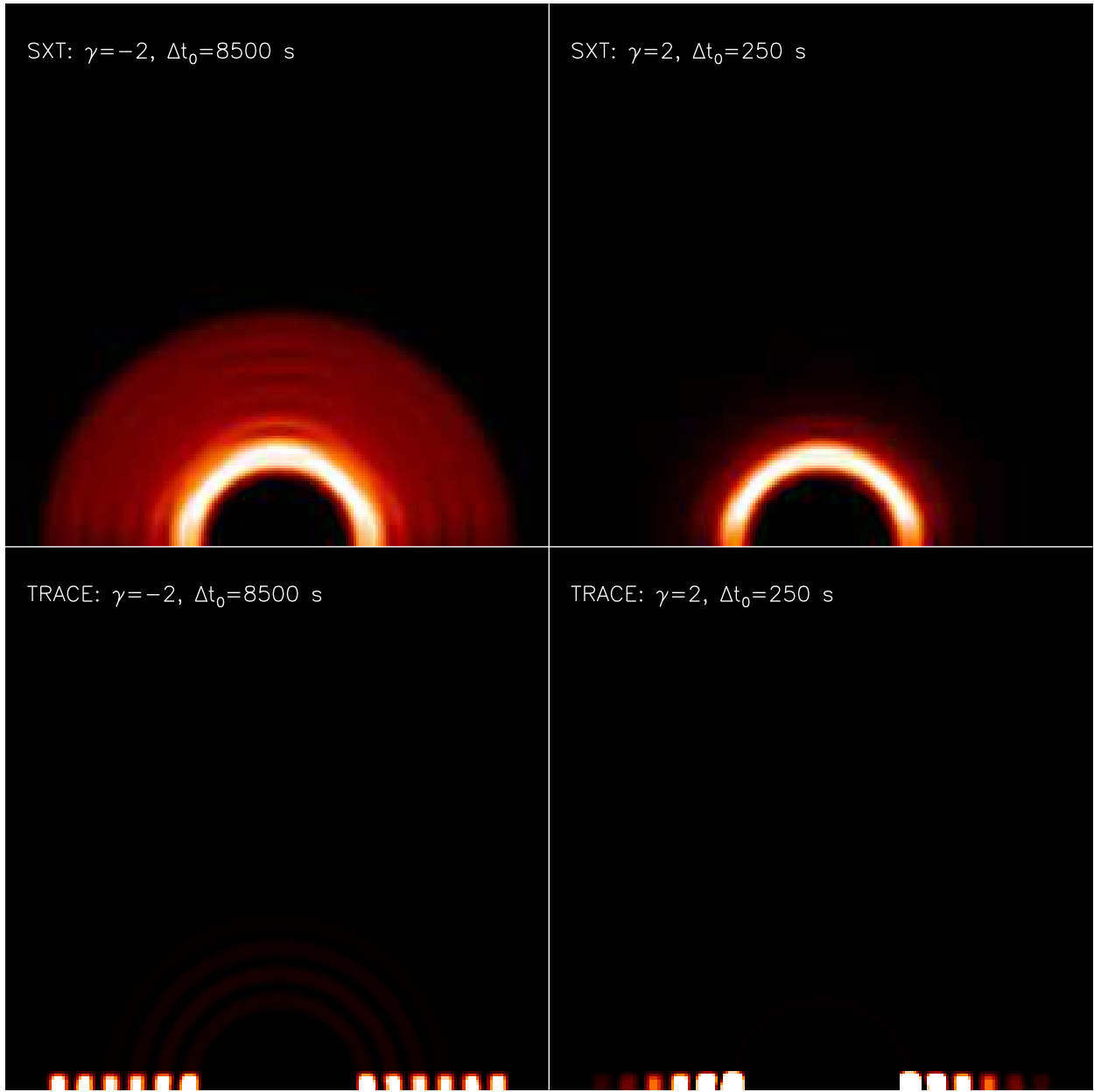}
\caption{Investigating the effect of the time-interval ${\Delta\tau}_{nano}$ between
successive nanoflares. Synthetic TRACE and SXT  images for   ${\Delta\tau}_{nano}$ which decreases
(increases) with loop length left (right) column. The AR baseline has  a length of $\approx$ 100 Mm. Intensity 
represented with color
increasing from black to red to white. Each image is normalized individually.}
\label{fig:im_gama}
\end{figure}

The impact  of the time-interval  between
successive nanoflares,  ${\Delta\tau}_{nano}$, assuming the dependence  
${\Delta\tau}_{nano}={\Delta\tau}_{0}{(L/L_{0})}^{\gamma}$ was then examined.  We consider two cases: 
(1) $\gamma=-2$ and ${\Delta\tau}_{0}=8500$ s; (2) $\gamma=2$ and ${\Delta\tau}_{0}=250$ s.
For each loop
we considered 3 consecutive nanoflares separated by ${\Delta\tau}_{nano}$.
The  nanoflare heating rate and duration were 
the same as in the base simulation which implies
the same averaged energy per nanoflare.
When  ${\Delta\tau}_{nano}$ decreases with $L$ (case 1)  the resulting images (left  column of Figure  \ref{fig:im_gama})
are characterized by  TRACE emission  concentrated
around the AR core  while the SXT emission is extended. 
This is because the longest loops in the arcade 
have ${\Delta\tau}_{nano}$ of about 850 s which is not
sufficient to let them
cool down to TRACE temperatures.
On the other hand, when
${\Delta\tau}_{nano}$ increases with $L$ (case 2), the resulting AR images are characterized by bright core emission
in SXT and lack of TRACE loops (right column of Figure  \ref{fig:im_gama}).
This is because shorter loops do not have the time to cool down
to TRACE temperatures and they remain in quasi-steady state conditions
in SXT temperatures, as also shown in Section 2.1. This model can explain observations
of an AR with SXT loops but no TRACE loops  (Antiochos et al. 2003).

\begin{figure}[!h]
\plotone{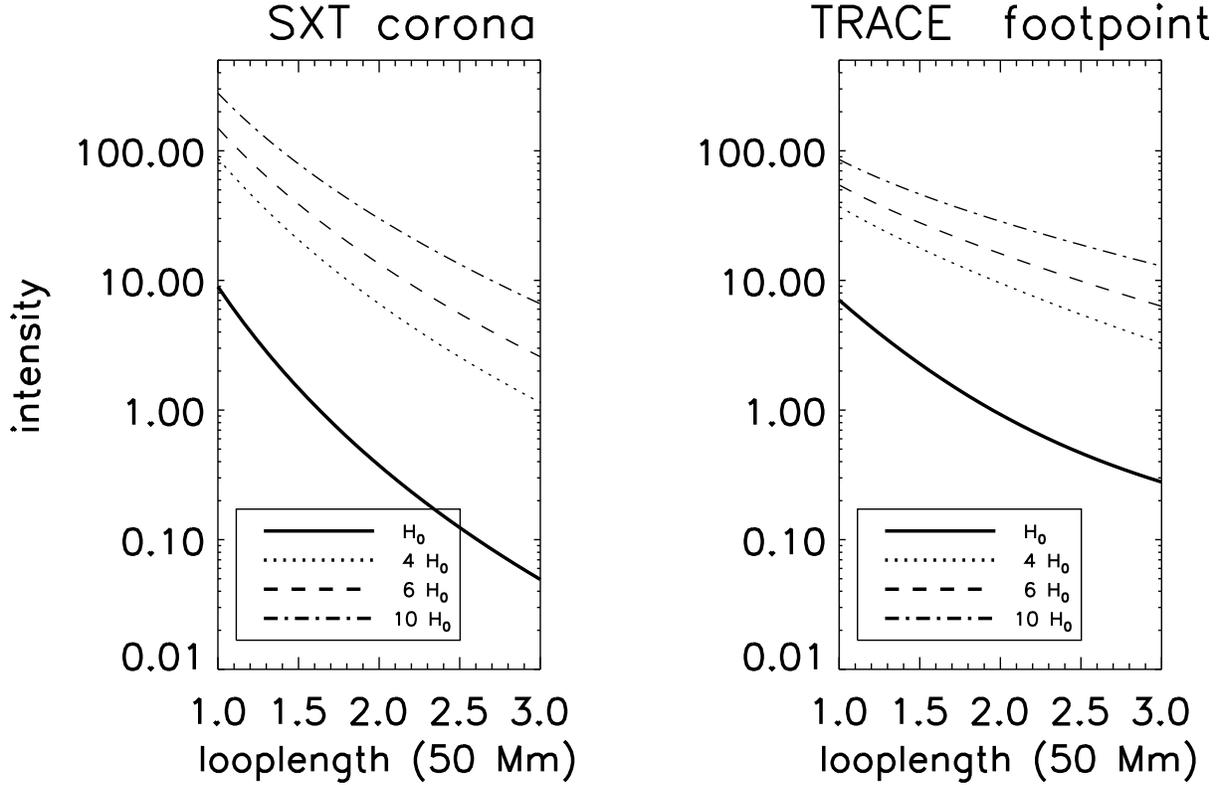}
\caption{Variation of the 
coronal and footpoint 
intensity across the loop arcade 
for SXT (left panel) and TRACE (right panel) respectively 
for impulsively heated AR models with different nanoflare energies.
The nanoflare magnitude for each AR model are 4 (dots), 6 (long dashes), 10 (dash-dot)
times that of the base model (solid).
The time-averaged intensities
are plotted. Intensities are in units of  DN/pix/s.}
\label{fig:im_e}
\end{figure}

We then investigated the effect of the magnitude of the nanoflare energy
on the SXT and TRACE coronal and moss intensities respectively.
These intensities are particularly useful when studying the cores
of active regions and can constrain 
the properties of the heating (e.g. Winebarger, Warren $\&$ Falconer 2007).
On top of the base  model described in the first paragraph of this section, 
we considered 3 additional AR models  employing  nanoflare heating magnitudes 2,6 and
10 times the one used in the base simulation (see also Equation 1). 
All other aspects of these simulations were the same as for the base simulation. 
The variation of the temporally averaged intensities across the loop arcade
for the 4 models is shown in Figure \ref{fig:im_e}.
What can be seen in this Figure is that increasing  nanoflare
energy leads to higher SXT coronal to TRACE moss intensity ratio  $I_{SXR}/I_{EUV-moss}$. 
For instance, for the inner  loops in the modeled AR (i.e. its core)
$I_{SXR}/I_{EUV-moss}$  
increases from $\approx$ 1.3 to 4.4 with a 10-fold
increase in the nanoflare energy. This means
that $I_{SXR}/I_{EUV-moss}$ tracks in a rather sensitive way
the nanoflare energy and can be used as its diagnostic.
Is it worth mentioning that the determined trend goes in the direction 
of decreasing the discrepancy with observations that 
can cause problems with static models.                                                                                                     
Possibly the spicular absorption discussed in the previous paragraphs could bring
$I_{SXR}/I_{EUV-moss}$ even to a better agreement with the observations

We finally found that the nanoflare duration does not
alter the AR morphology for  both SXT and TRACE.
This comes to no surprise given
both instruments are mostly sensitive to the late cooling
of the impulsively heated loops, when any differences
in the loop response for different nanoflare durations,
had been long ago smeared out (e.g. Patsourakos $\&$ Klimchuk 2005).

\section{Conclusion}
With this work we considered static and
impulsive heating models of active region arcades. In concert with previous investigations, 
we found that static models  cannot simultaneously reproduce
bright EUV and SXR loops withing the same AR. We showed that this is due to the shallow
dependence of loop temperature on loop length implied by all reasonable coronal 
heating scenarios.
We found that impulsive heating models agree much better with observations than 
do static models.  In particular, impulsive heating produces (1) a reduced brightness 
contrast between the corona and footpoints in TRACE observations, (2) an increased 
TRACE-to-SXT coronal intensity ratio, and (3) enhanced SXT emission in the core 
of the active region and extended TRACE emission. 
We finally showed that
the AR morphology depends rather sensitively on the properties
of impulsive heating, like its spatial dependence and the time interval between
successive nanoflares, and can therefore be used to determine the properties
of the heating. Our study paves the way for detailed comparisons between
multi-temperature observations of ARs and models based on impulsive heating
and detailed reconstructions of the coronal magnetic field from extrapolations.

Research supported by NASA and ONR. We acknowledge 
useful discussions with the members of the ISSI team "The role
of Spectroscopic and Imaging Data in Understanding Coronal Heating" (team
Parenti).



\clearpage

\clearpage

\clearpage

\clearpage

\clearpage

\clearpage

\end{document}